\documentstyle[12pt]{article}

\begin{document}
\title{ Perturbed Coulomb potentials in the Klein - Gordon equation via
the shifted - $l$ expansion technique}
\author{ Thabit Barakat \\
 Department of Civil Engineering, Near East University\\
 Lefko{\c s}a, North Cyprus, Mersin 10 - Turkey\\
 Maen Odeh and Omar Mustafa$^{\dag}$ \\
 Department of Physics, Eastern Mediterranean University\\
 G. Magusa, North Cyprus, Mersin 10 - Turkey\\
\date{}\\}
\maketitle
\begin{abstract}
{\small A shifted - $l$ expansion technique is introduced to
calculate the energy eigenvalues for Klein - Gordon (KG) equation with 
Lorentz vector and/or Lorentz scalar potentials. Although it applies to any
spherically symmetric potential, those that include Coulomb -
like terms are only considered. Exact eigenvalues for a Lorentz
vector or a Lorentz scalar, and an equally mixed Lorentz vector and Lorentz 
scalar Coulombic potentials are reproduced. Highly accurate and rapidly
converging  ground - state energies for  Lorentz vector Coulomb with a 
Lorentz vector or a Lorentz scalar linear potential, $V(r)=-A_{1}/r+kr$, and
$V(r)=-A_{1}/r$ and $S(r)=kr$, respectively,
 are obtained. Moreover, a simple
straightforward closed - form solution for a KG - particle in
Coulombic Lorentz vector and Lorentz scalar potentials is presented in 
appendix A.}
\end{abstract}
\newpage
\renewcommand{\thesection}{\Roman{section}}

\section{Introduction}

The Klein - Gordon ( KG) and the Dirac equations with Lorentz scalar 
( added to the mass term) and/or Lorentz vector ( coupled as the 
0 - component of the four - vector potential )
potentials are of interest in many branches of physics. For example, 
Lorentz scalar or  equally mixed Lorentz scalar and Lorentz vector 
potentials have
considerable interest in quark - antiquark mass spectroscopy [1-5].
Lorentz vector potentials have great utility in atomic, nuclear, and
plasma physics [6,7]. Therefore many attempts have been made to
develop approximation techniques to treat relativistic particles in the 
KG and the Dirac equations [1-7].

Very recently we have introduced a shifted - $l$ expansion technique
( SLET ) to solve for the Schr\"odinger [8], 
and the Dirac equations for some model potentials [9].
SLET is a reformation to the existing
shifted - N expansion technique ( SLNT ) [1,10-12, and references
therein]. SLET simply consists of using $1/\bar{l}$ as an expansion
parameter where $\bar{l} = l - \beta$, $\beta$ is a suitable shift,
$l$ is the angular momentum quantum number for spherically symmetric
potentials, and $l = |m|$ for cylindrically symmetric potentials,
where m is the magnetic quantum number. As such, one does not need to
construct the N - dimensional form, required to perform SLNT, of the
wave equation of interest. With SLET we simply expand 
through the quantum number 
in the centrifugal term of that equation. Unlike other perturbation
methods [13-16], SLET puts no constraints on the coupling constants of
the potential or on the quantum numbers involved. Above all, it yields
very accurate and rapidly converging eigenvalues without the need of
wave functions  or matrix elements.

In this paper we shall be concerned with the shifted - $l$ expansion
for the KG equation with radially symmetric Lorentz scalar, $S(r)$, and/or 
Lorentz vector, $V(r)$, potentials that include Coulomb - like terms. 
We shall
examine SLET and calculate the energy eigenvalues for the KG equation
with the following potential mixtures: (i) $V(r)=-A_{1}/r$ and
$S(r)=0$, which represents a $\pi^{-}$ meson in a Coulomb
potential. (ii) $V(r)=0$ and $S(r)=-A_{2}/r$, which has no experimental
evidence, to the best of our knowledge, thus our calculations are only
of academic interest. (iii)  $V(r)=S(r)=-A/r$ which represents not
only a KG - particle in an equally mixed Lorentz scalar and Lorentz vector
potential but also a Dirac particle in the same potential mixture,
where $l=j+1/2$ and the radial KG wave function represents the radial
large - component of the Dirac - spinor [4,9]. (iv) $V(r)=-A_{1}/r+kr$
and $S(r)=0$ representing a $\pi^{-}$ meson in a Coulomb potential
perturbed by a linear Lorentz vector interaction $kr$. (v) $V(r)=-A_{1}/r$
and $S(r)=kr$ describing a $\pi^{-}$ meson in a Coulomb potential
perturbed by a linear Lorentz scalar potential $kr$.

In Sec II we shall introduce SLET for the KG equation with any
spherically symmetric Lorentz scalar and/or Lorentz vector potentials 
that include
Coulomb - like interactions. We shall cast SLET's analytical
expressions in such a way that allows the reader to use them without
proceeding into their derivations. In Sec III we shall show that these
expressions yield closed - form solutions to the KG equation for the
mixtures (i), (ii), and (iii). Ground - state energies for the
mixtures (iv) and (v) will be calculated and compared with those of
McQuarrie and Vrscay [13] in the same section. We conclude and remark
in Sec IV.

In appendix A we present a simple straightforward closed - form 
solution for the KG
equation with Coulomb - like Lorentz scalar and Lorentz vector potentials.
It could be interesting to mention that a similar solution was found by
McQuarrie and Vrscay [13], who used a confluent hypergeometric
function in their calculation. They have misprinted it though ( see
appendix of Ref [13]). To the best of our knowledge, such explicit
solution has not been reported elsewhere.

\section{ SLET for the KG equation with potentials including Coulombic
terms.}

In this section we shall consider the three - dimensional KG equation
with radially symmetric Lorentz vector and Lorentz scalar potentials, 
$V(r)$ and
$S(r)$, respectively. If $\Psi({\bf r})$ denotes the wave function of the KG
particle, a separation of variables $\Psi(r)=r^{-1}R(r)Y(\theta,\phi)$
yields the following radial equation ( in units $\hbar=c=1$) [1]:
\begin{equation}
\left[-\frac{d^{2}}{dr^{2}}+\frac{l(l+1)}{r^{2}}+[S(r)+m]^{2}
-[E-V(r)]^{2}\right]R(r)=0,
\end{equation}
where $E$ is the energy, and $l$ is the angular quantum number. For
Coulomb - like potentials one may use the substitutions:
\begin{equation}
V_{r}(r)=V(r)^{2}-A_{1}^{2}/r^{2},
\end{equation}
and\\
\begin{equation}
S_{r}(r)=S(r)^{2}-A_{2}^{2}/r^{2},
\end{equation}
so that Eq.(1) becomes\\
\begin{equation}
\left[-\frac{d^{2}}{dr^{2}}+\frac{l^{'}(l^{'}+1)}{r^{2}}+
\gamma(r)+2EV(r)\right]R(r)=E^{2}R(r),
\end{equation}
where\\
\begin{equation}
\gamma(r)=-V_{r}(r)+S_{r}(r)+2mS(r)+m^{2},
\end{equation}
\begin{equation}
l^{'}(l^{'}+1)=l(l+1)-A_{c};{~~}l^{'}=-1/2+\sqrt{(l+1/2)^{2}-A_{c}},
\end{equation}
\begin{equation}
A_{c}=A_{1}^{2}-A_{2}^{2}.
\end{equation}
Hereby, it should be noted that for the case of $V(r)=-A_{1}/r$ and
$S(r)=-A_{2}/r$ Eq.(4) reduces to a form nearly identical to the
Schr\"odinger equation for a Coulomb field. Its solution can thus 
be inferred from the known solution of the 
Schr\"odinger - Coulomb problem. We do this in appendix A.

If we shift $l^{'}$ through the relation
$l^{'}=\bar{l}+\beta$ Eq.(4) reads\\
\begin{equation}
\left[-\frac{d^{2}}{dr^{2}}+\frac{[\bar{l}^{2}+\bar{l}(2\beta+1)+
\beta(\beta +1)]}{r^{2}}+\gamma(r)+2EV(r)\right]R(r)=E^{2}R(r).
\end{equation}
where $\beta$ is a suitable shift to be determined and is mainly
introduced to avoid the trivial case when $l^{'}=0$.

To start the systematic $1/\bar{l}$  expansion [8,9] we define\\
\begin{equation}
\gamma(r)=\frac{\bar{l}^{2}}{Q}\left[\gamma(r_{o})+\gamma'(r_{o})r_{o}
 x/\bar{l}^{1/2}
+\gamma''(r_{o})r_{o}^{2}x^{2}/2\bar{l}+\cdots\right],
\end{equation}
\begin{equation}
V(r)=\frac{\bar{l}^{2}}{Q}\left[V(r_{o})+V'(r_{o})r_{o}
 x/\bar{l}^{1/2}
+V''(r_{o})r_{o}^{2}x^{2}/2\bar{l}+\cdots\right],
\end{equation}
\begin{equation}
E=\frac{\bar{l}^{2}}{Q}\left[E_{o}+E_{1}/\bar{l}+E_{2}/
 \bar{l}^{2}+E_{3}/\bar{l}^{3}+\cdots\right].
\end{equation}
where $x=\bar{l}^{1/2}(r-r_{o})/r_{o}$, $r_{o}$ is currently an arbitrary
point to do Taylor expansions about, with its particular value to be
determined below,  and $Q$ is to be set equal
to $\bar{l}^{2}$ at the end of the calculations. Substituting 
Eqs (9) - (11) into Eq.(8) implies\\
\begin{eqnarray}
&&\left[\frac{-d^{2}}{dx^{2}}+(\bar{l}+(2\beta+1)+\frac{\beta(\beta+1)}
{\bar{l}})(1-\frac{2x}{\bar{l}^{1/2}}+\frac{3x^{2}}{\bar{l}}
-\cdots)\right. \nonumber \\
&&\nonumber\\
&&\left.+\frac{r_{o}^{2}\bar{l}}{Q}(\gamma(r_{o})+\frac{\gamma'(r_{o})r_{o}
x}{\bar{l}^{1/2}}+\frac{\gamma''(r_{o})r_{o}^{2}x^{2}}{2\bar{l}}+
\frac{\gamma'''(r_{o})r_{o}^{3}x^{3}}{6\bar{l}^{3/2}}+\cdots)\right.
\nonumber \\
&&\nonumber \\
&&\left.+\frac{2r_{o}^{2}\bar{l}}{Q}(V(r_{o})+\frac{V'(r_{o})r_{o}x }
{\bar{l}^{1/2}} +\cdots)(E_{o}+\frac{E_{1}}{\bar{l}}
+\frac{E_{2}}{\bar{l}^{2}}+\cdots)\right]\Phi_{n_{r}}(x)
\nonumber \\
&&\nonumber \\
&&=\mu_{n_{r}}\Phi_{n_{r}}(x)
\end{eqnarray}\\
 where\\
\begin{equation}
\mu_{n_{r}}=\frac{r^{2}_{0}\bar{l}}{Q}\left[
E_{o}^{2}+\frac{2E_{o}E_{1}}{\bar{l}}
+\frac{(E_{1}^{2}+2E_{o}E_{2})}{\bar{l}^{2}}+\frac{
2(E_{o}E_{3}+E_{1}E_{2})}{\bar{l}^{3}}
+\cdots\right],
\end{equation}
and $n_{r}$ is the radial quantum number.
Eq.(12) is a Schr\"odinger-like equation for the one-dimensional anharmonic
oscillator and has been discussed in detail by Imbo et al [11]. We
therefore quote only the resulting eigenvalue of Ref.[11] and write\\
\begin{eqnarray}
\mu_{n_{r}}&=&\bar{l}\left[1+\frac{2r_{o}^{2}V(r_{o})E_{o}}{Q}+
\frac{r_{o}^{2}\gamma(r_{o})}{Q} \right]
\nonumber \\
&&\nonumber \\
&&+\left[(2\beta+1)+\frac{2r_{o}^{2}V(r_{o})E_{1}}{Q}+(n_{r}+\frac{1}{2})w
\right]
\nonumber \\
&&\nonumber\\
&&+\frac{1}{\bar{l}}\left[\beta(\beta+1)+\frac{2r_{o}^{2}V(r_{o})E_{2}}{Q}
+\alpha_{1}\right]
\nonumber \\
&&\nonumber\\
&&+\frac{1}{\bar{l}^{2}}\left[\frac{2r_{o}^{2}V(r_{o})E_{3}}{Q}+\alpha_{2}
\right],
\end{eqnarray}\\
where $\alpha_{1}$  and $\alpha_{2}$  are given in appendix B of
this text. If we compare Eq.(14) with (13), we obtain\\
\begin{equation}
E_{o}=V(r_{o})\pm\sqrt{V(r_{o})^{2}+Q/r_{o}^{2}+\gamma(r_{o})},
\end{equation}
\begin{equation}
E_{1}=\frac{Q}{2r_{o}^{2}(E_{o}-V(r_{o}))}\left[2\beta+1+(n_{r}+1/2)w\right],
\end{equation}
\begin{equation}
E_{2}=\frac{Q}{2r_{o}^{2}(E_{o}-V(r_{o}))}\left[\beta(\beta+1)+\alpha_{1}
\right],
\end{equation}
\begin{equation}
E_{3}=\frac{Q}{2r_{o}^{2}(E_{o}-V(r_{o}))}\alpha_{2},
\end{equation}
and\\
\begin{equation}
E_{n_{r}}=E_{o}+\frac{1}{2r_{o}^{2}(E_{o}-V(r_{o}))}\left[\beta(\beta+1)+
\alpha_{1}+\frac{\alpha_{2}}{\bar{l}}\right].
\end{equation}
$r_{o}$ is chosen to be the  minimum of $E_{o}$, i. e.;\\
\begin{equation}
dE_{o}/dr_{o}=0 ~~~~and~~~~d^{2}E_{o}/dr_{o}^{2}>0.
\end{equation} \\
 Hence, $r_{o}$  is obtained through the relation\\
\begin{equation}
2Q=2(l^{'}-\beta)^{2}=b(r_{o})+\sqrt{b(r_{o})^{2}-4c(r_{o})},
\end{equation}
where\\
\begin{equation}
b(r_{o})=r_{o}^{3}\left[2V(r_{o})V^{'}(r_{o})+\gamma^{'}(r_{o})+r_{o}V^{'}
(r_{o})^{2}\right],
\end{equation}
\begin{equation}
c(r_{o})=\frac{r_{o}^{6}}{4}\left[\gamma^{'}(r_{o})^{2}+4V(r_{o})V^{'}(r_{o})
\gamma^{'}(r_{o})-4\gamma(r_{o})V^{'}(r_{o})^{2}\right]
\end{equation}
The shifting parameter $\beta$ is determined by requiring
$E_{1}=0$ [1, 8-12] to obtain\\
\begin{equation}
\beta=-[1+(n_{r}+1/2)w]/2,
\end{equation}
where\\
\begin{equation}
w=\left[12+\frac{2r_{o}^{4}\gamma^{''}(r_{o})}{Q}+
\frac{4r_{o}^{4}V^{''}(r_{o})E_{o}}{Q}\right]^{1/2}.
\end{equation}
It is convenient to summarize the above procedure in the following
steps: (a) Calculate $Q$ from Eq.(21) and substitute it in Eq.(15) to
find $E_{o}$ in terms of $r_{o}$. (b) Substitute $E_{o}$ and $Q$ in
Eq.(25) to obtain $w$. (c) Find $\beta$ from Eq.(24) to calculate
$r_{o}$ from Eq.(21). (e) Finally, one can obtain $E_{o}$ and
calculate $E_{n_{r}}$ from Eq.(19). However, one is not always able to
calculate $r_{o}$ in terms of the potential coupling constants since
the analytical expressions become algebraically complicated, although
straightforward. Therefore, one has to appeal to numerical computations
to find $r_{o}$ and hence $E_{o}$.

\section{ Applications, results, and discussion.}

To show the performance of the analytical expressions of SLET it is
best to consider some special cases.

(i) $V(r)=-A_{1}/r$ and $S(r)=0$

A pionic atom in a Coulomb potential
obeys the KG equation with $V(r)=-A_{1}/r$ and
$S(r)=0$. To calculate its bound - state energies, which are simply
the bound state energies of a $\pi^{-}$ meson in a Coulomb potential, we
follow the SLET procedure and find\\
\begin{equation}
E_{o}=\frac{-A_{1}^{2}\pm(A_{1}^{2}+Q)}{A_{1}r_{o}}.
\end{equation}
Here we have to choose the positive sign since states with negative 
energies correspond to anti - particles. Furthermore, the negative sign 
yields a contradiction to Eq.(21). Hence $w=2$,\\
\begin{equation}
Q=(l^{'}-\beta)^{2}=[n_{r}+1/2+\sqrt{(l+1/2)^{2}-A_{1}^{2}}]^{2}.
\end{equation}
\begin{equation}
r_{o}=\sqrt{\frac{Q^{2}+QA_{1}^{2}}{m^{2}A_{1}^{2}}},
\end{equation}
and\\
\begin{equation}
E_{o}=m[1+\frac{A_{1}^{2}}{\tilde{n}^{2}}]^{-1/2},
\end{equation}
where $\tilde{n}=\sqrt{Q}$. Eq.(29) represents the
well known closed - form solution of the KG equation for a $\pi^{-}$
meson in a Coulomb potential [17]. It should be noted that higher -
order terms of the energy eigenvalues vanish identically, i. e.
$E_{2}=0$ and $E_{3}=0$. Hence $E_{n_{r}}=E_{o}$.

(ii) $V(r)=0$ and $S(r)=-A_{2}/r$

Since there is no experimental evidence, to the best of our
knowledge, for such a long - range interaction, our
calculations are only of academic interest. Following the above procedure
we find\\
\begin{equation}
r_{o}=\frac{(l^{'}-\beta)^{2}}{mA_{2}},
\end{equation}
and\\
\begin{equation}
E_{o}=\pm m[1-\frac{A_{2}^{2}}{\tilde{n}^{2}}]^{1/2},
\end{equation}
where $\tilde{n}=n_{r}+1/2+\sqrt{(l+1/2)^{2}+A_{2}^{2}}$.
Again the higher - order terms of the energy eigenvalues vanish
identically. Thus $E_{n_{r}}=E_{o}$.

Obviously there exist two branches of solutions in the bound region
and they exhibit identical behaviour, which reflects the fact that the
Lorentz scalar interaction does not distinguish between particle and
antiparticle. The particle and antiparticle states, positive and
negative energies, respectively, approach each other with increasing
coupling constant, without touching. Therefore, spontaneous pair
creation never occurs, no matter how strong the potential chosen. 

(iii) $V(r)=S(r)=-A/r$

This type of potential mixture, $V(r)=S(r)$, has considerable
interest in quarkonium spectroscopy [4,9,18]. For the particular case
$V(r)=S(r)=-A/r$, SLET yields\\
\begin{equation}
E_{o}=-A/r_{o}\pm m,
\end{equation}
and \\
\begin{equation}
(E_{o}^{2}-m^{2})r_{o}^{2}=A^{2}\mp(Q+A^{2}).
\end{equation}
Eq.(33) can be satisfied if and only if the negative sign is chosen,
otherwise it contradicts Eq.(21). The only valid sign in Eq.(32) is thus
the positive one, and hence\\
\begin{equation}
E_{o}=-A/r_{o}+m.
\end{equation}
Which in turn implies that\\
\begin{equation}
r_{o}=\frac{n^{2}+A^{2}}{2mA};{~~}n=n_{r}+l+1,
\end{equation}
and \\
\begin{equation}
E_{n_{r}}=E_{o}=m[1-\frac{2A^{2}}{n^{2}+A^{2}}].
\end{equation}
Where higher - order terms of the energy eigenvalues vanish
identically, and n is the principle quantum number. For
$A\rightarrow\infty$, $E_{n_{r}}$ approaches the value $-m$
asymptotically, but the state never dives into the negative continuum.

To show that Eq.(36) yields the energy eigenvalues for Dirac particle
in the same potential mixture, we replace $l$ by $j+1/2$ to obtain\\
\begin{equation}
E_{n_{r}}=m[1-\frac{2A^{2}}{(n_{r}+|\kappa|+1)^{2}+A^{2}}].
\end{equation}
where $|\kappa|=j+1/2$ [19].

(iv) $V(r)=-A_{1}/r+kr$ and $S(r)=0$

This potential represents a $\pi^{-}$ meson in a Coulomb potential
perturbed by a linear Lorentz vector potential $kr$. In this case \\
\begin{equation}
\gamma(r)=-k^{2}r^{2}+2A_{1}k+m^{2}.
\end{equation}
Eq.(38) when substituted in (21), (15), (25), (24), and again in (21),
respectively, yields a very involved algebraic equation for $r_{o}$.
We solve this equation numerically with a maximum error of
order $\sim 10^{-15}$ to calculate for $r_{o}$. Once $r_{o}$ is
calculated, $Q$, $E_{o}$, $w$, $\beta$, and hence $E_{n_{r}}$ can be
obtained.

In Tables 1 and 2 we list our results for the ground - state energies along
with those of McQuarrie and Vrscay [13], who have used hypervirial and
Hellmann - Feynman theorems to construct Rayleigh - Schr\"odinger (RS)
perturbation expressions to an arbitrary order. Our results are given
in such a way that the contributions of the second - and third - order
corrections, $E_{2}/\bar{l}^{2}$ and $E_{3}/\bar{l}^{3}$,
respectively, to the energy eigenvalues are made clear. The results
are in excellent agreement with those of Ref.[13].

(v) $V(r)=-A_{1}/r$ and $S(r)=kr$

A $\pi^{-}$ meson in a Coulomb potential perturbed by a linear scalar
interaction is described by $V(r)=-A_{1}/r$ and $S(r)=kr$ potential
mixture in the KG equation. In this case\\
\begin{equation}
\gamma(r)=k^{2}r^{2}+2mkr+m^{2}.
\end{equation}
Following the same steps of (iv) we numerically solve for
$r_{o}$, to a maximum error of order $\sim 10^{-15}$, $Q$, $E_{o}$,
$w$, $\beta$, and $E_{n_{r}}$. Our results for the ground - state
energies are presented in Tables 3 and 4 in such a way that the convergence 
of SLET is made clear. We compare them with those of Ref.[13]. They are
in excellent agreement.

In view of the above results the following observations  deserve to be
recorded.

The closed - form solutions, Eqs.(29), (31), and (36), being obtained
by the leading term of the energy series, Eq.(11), where higher -
order terms vanished identically, reveals how rapidly converging are
the results of SLET.

The numerical results of SLET, in Tables 1 - 4, imply that the
contributions of the second - and third - order corrections to the
energy eigenvalues are almost negligible. The convergence of SLET is
thus out of question. However, the RS coefficients $E^{p}$ for the
eigenvalue\\
\begin{equation}
E=\sum_{p=0}^{\infty}E^{(p)}k^{p},
\end{equation}
used in Ref.[13], as well as their continued - fraction (CF)
counterparts $c_{p}$ were computed numerically to large - order,
$p\sim100$ and $p\sim50$ for the Lorentz vector linear and the Lorentz scalar
linear perturbations, respectively. Numerical ratio tests showed that
the perturbation series are divergent [13]. As well, the $c_{p}$ had
suffered from occasional eruptions reversing the roles of the upper
and lower bounds of the energy eigenvalues. Moreover, the gap between
the two bounds increases with increasing coupling constant $k$, same is
the uncertainty of the energy eigenvalues.
    
\section{ Conclusions and remarks}

In this paper we have introduced SLET to solve for the eigenvalues of
KG equation with Lorentz vector and Lorentz scalar potentials including 
Coulombic
terms. Although it applies to any spherically symmetric potential,
those that include Coulomb - like terms were only considered. We have
reproduced closed - form solutions for a Lorentz vector or a Lorentz 
scalar, and
for an equally mixed Lorentz vector and Lorentz scalar Coulombic 
potentials [20].
Compared to those of Ref.[13] our results are highly accurate and
rapidly convergent.

The conceptual soundness of our SLET is obvious. It is highly accurate 
and efficient with respect to computers time. It does not need the wave
functions or matrix elements, but when necessary
wave functions can be calculated.
It puts no constraints on the coupling constants of the
potential or on the quantum numbers. It simply consists of using
$1/\bar{l}$ as an expansion parameter rather than the coupling
constant of the potential. It is to be understood as being an
expansion through not only the angular momentum quantum number but
also through any existing quantum number in the centrifugal - like term
of any Schr\"odinger - like equation, Eq.(4).

A general observation concerning the method used by McQuarrie and
Vrscay [13] is in order. Unlike our approach their method involves
expansions through the coupling constant $k$, Eq.(40). Thus, whereas
their computations for the ground - state energies are beyond doubt,
the same need not be true for the case of strong coupling constant
$k>1$, in Eq.(40), for example.

Finally, we would like to remark that  SLET is also
applicable to more complicated potentials. For example, the screened
Coulomb potentials which have great utility in atomic, nuclear, and
plasma physics. The equally mixed Lorentz scalar and Lorentz vector 
logarithmic
potential which has significant interest in quarkonium spectroscopy
[4].

\newpage
\begin{center}
{\bf\Large{Appendix A}\\ \large{The KG equation with Coulomb - like
Lorentz scalar and Lorentz vector potentials}}
\end{center}

In this section we present a simple solution for a KG particle in
Coulomb - like Lorentz scalar and Lorentz vector potentials, 
i. e. $V(r)=-A_{1}/r$
and $S(r)=-A_{2}/r$. For this particular problem the KG equation 
reduces to\\
\begin{equation}
\left[-\frac{d^{2}}{dr^{2}}+\frac{l^{'}(l^{'}+1)}{r^{2}}-
\frac{2(mA_{2}+EA_{1})}{r} \right]R(r)=[E^{2}-m^{2}]R(r).
\end{equation}
It is obvious that this equation is in a form nearly identical to the
Schr\"odinger equation for a Coulomb potential.
Its solution can thus be inferred
from the known Schr\"odinger - Coulomb solution. Therefore, one may
obtain its solution through the relation\\
\begin{equation}
E^{2}-m^{2}=\frac{-(2mA_{2}+2EA_{1})^{2}}{(2\tilde{n})^{2}}.
\end{equation}
This equation is quadratic in $E$ and thus admits a solution of the
form\\
\begin{equation}
E_{n_{r}}=m\left[\frac{-A_{1}A_{2}\pm
\sqrt{A_{1}^{2}A_{2}^{2}+(\tilde{n}^{2}+A_{1}^{2})(\tilde{n}^{2}-A_{2}^{2})}}
{\tilde{n}^{2}+A_{1}^{2}}\right],
\end{equation}
where $\tilde{n}=n_{r}+l^{'}+1$. Hereby, it should be pointed out that
this result reduces to those obtained in Sec.III.1,2, and 3. Although
McQuarrie and Vrscay [13] have used a confluent hypergeometric function 
to obtain this result, they have misprinted it (see the appendix of
Ref.[13]).
\newpage
\begin{center}
{\bf\Large{Appendix B}\\ \large{$\alpha_{1}$ and $\alpha_{2}$ in Eq.(14)} }
\end{center}
The definitions of $\alpha_{1}$ and $\alpha_{2}$ appeared in Eq.(14)
are:\\
\begin{eqnarray}
\alpha_{1}&=&[(1+2n_{r})e_{2}+3(1+2n_{r}+2n_{r}^{2})e_{4}]\nonumber\\
&&\nonumber\\
&-&w^{-1}[e_{1}^{2}+6(1+2n_{r})e_{1}e_{3}+(11+30n_{r}+30n_{r}^{2})
e_{3}^{2}],
\end{eqnarray}
\begin{eqnarray}
\alpha_{2}&=& (1+2n_{r})d_{2}+3(1+2n_{r}+2n_{r}^{2})d_{4}\nonumber\\
&&\nonumber\\
&+&5(3+8n_{r}+6n_{r}^{2}+4n_{r}^{3})d_{6}\nonumber\\
&&\nonumber\\
&-&w^{-1}[(1+2n_{r})e_{2}^{2}
+12(1+2n_{r}+2n_{r}^{2})e_{2}e_{4}
+2e_{1}d_{1}
\nonumber\\
&&\nonumber\\
 &+&2(21+59n_{r}+51n_{r}^{2}+34n_{r}^{3})e_{4}^{2}
+6(1+2n_{r})e_{1}
d_{3}\nonumber\\
&&\nonumber\\
 &+&30(1+2n_{r}+2n_{r}^{2})e_{1}d_{5}
+6(1+2n_{r})e_{3}d_{1}\nonumber\\
&&\nonumber\\
&+&2(11+30n_{r}+30n_{r}^{2})e_{3}d_{3}+
10(13+40n_{r}+42n_{r}^{2}+28n_{r}^{3})e_{3}d_{5}]\nonumber\\
&&\nonumber\\
&+&w^{-2}[4e_{1}^{2}e_{2}
+36(1+2n_{r})e_{1}e_{2}
e_{3}+8(11+30n_{r}
+30n_{r}^{2})e_{2}e_{3}^{2}\nonumber\\
&&\nonumber\\
&+&24(1+n_{r})e_{1}^{2}e_{4}
+8(31+78n_{r}+78n_{r}^{2})e_{1}e_{3}
e_{4}\nonumber\\
&&\nonumber\\
&+&12(57+189n_{r}+225n_{r}^{2}+150n_{r}^{3})
e_{3}^{2}e_{4}]
\nonumber\\
&&\nonumber\\
&-&w^{-3}[8e_{1}^{3}e_{3}
+108(1+2n_{r})e_{1}^{2}e_{3}^{2}
+48(11+30n_{r}+30n_{r}^{2})e_{1}e_{3}^{3}
\nonumber\\
&&\nonumber\\
&+&30(31+109n_{r}+141n_{r}^{2}+94n_{r}^{3})e_{3}^{4}],
\end{eqnarray}
where\\
\begin{equation}
e_{j}=\frac{\varepsilon_{j}}{w^{j/2}}  \mbox{~~ and ~~}
d_{i}=\frac{\delta_{i}}{w^{i/2}}\,,
\end{equation}
with $j=1,2,3,4$, $i=1,2,3,4,5,6$, and\\
\begin{equation}
\varepsilon_{1}=-2(2\beta+1) {~~,~~}
\varepsilon_{2}=3(2\beta+1),
\end{equation}
\begin{equation}
\varepsilon_{3}=-4+\frac{r_{o}^{5}}{6Q}\left[\gamma^{'''}(r_{o})+2V^{'''}
(r_{o})E_{o}\right],
\end{equation}
\begin{equation}
\varepsilon_{4}=5+\frac{r_{o}^{6}}{24Q}\left[\gamma^{''''}(r_{o})+2V^{''''}
(r_{o})E_{o}\right],
\end{equation}
\begin{equation}
\delta_{1}=-2\beta(\beta+1)+\frac{2r_{o}^{3}V^{'}(r_{o})E_{2}}{Q}
\end{equation}
\begin{equation}
\delta_{2}=3\beta(\beta+1)+\frac{r_{o}^{4}V^{''}(r_{o})E_{2}}{Q},
\end{equation}
\begin{equation}
\delta_{3}=-4(2\beta+1) {~~,~~}
\delta_{4}=5(2\beta+1),
\end{equation}
\begin{equation}
\delta_{5}=-6+\frac{r_{o}^{7}}{120Q}\left[\gamma^{'''''}(r_{o})+2V^{'''''}
(r_{o})E_{o}\right],
\end{equation}
\begin{equation}
\delta_{6}=7+\frac{r_{o}^{8}}{720Q}\left[\gamma^{''''''}(r_{o})+2V^{''''''}
(r_{o})E_{o}\right].
\end{equation}
The terms including $E_{1}$ have been
dropped from the expressions above since $E_{1}=0$.

\newpage
\begin{table}
\begin{center}
\caption{Ground - state energies of a $\pi^{-}$ meson in
$V(r)=-A_{1}/r+kr$ and $S(r)=0$ (in $\hbar=c=m=1$ units).
The lower bounds to the
energies $E$ of Ref.[13] are obtained by replacing the last $j$ digits
of the upper bounds with the $j$ digits in parentheses.}
\begin{tabular}{|llllll|}
\hline
$A_{1}$ & $k$ & Ref.[13] & $E_{o}$ & $E_{o}+E_{2}/\bar{l}^{2}$ &
Eq.(11) \\
 \hline
0.2&0.0&0.97890631293&0.97890631293&0.97890631293&0.97890631293\\
   & 0.01 & 1.027622(19) & 1.029590 & 1.027995 & 1.027641 \\
   & 0.05 & 1.152(48) & 1.1541 & 1.1514 & 1.1504 \\
   & 0.1  & 1.277(48) & 1.2681 & 1.2648 & 1.2634 \\
   & 0.2  & 1.50(37)  & 1.4478 & 1.4436 & 1.4416 \\
   & 0.3  & 1.73(45)  & 1.5959 & 1.5907 & 1.5882 \\
0.3&0.0&0.9486832981&0.9486832981&0.9486832981&0.9486832981\\
   &0.01&0.9843795380(78)&0.9861391713&0.9844578890&0.9843836237\\
   &0.05&1.08612(08) & 1.09160 & 1.08710 & 1.08611\\
   &0.1 &1.18398(08) & 1.19111 & 1.18517 & 1.18356\\
   &0.2 &1.345(34) & 1.3500 & 1.3421 & 1.3397\\
   &0.3 & 1.487(52)& 1.4820 & 1.4723 & 1.4693\\
\hline
\end{tabular}
\end{center}
\end{table}
\newpage
\begin{table}
\begin{center}
\caption{Ground - state energies of a $\pi^{-}$ meson in
$V(r)=-A_{1}/r+kr$ and $S(r)=0$ (in $\hbar=c=m=1$ units).
The lower bounds to the
energies $E$ of Ref.[13] are obtained by replacing the last $j$ digits
of the upper bounds with the $j$ digits in parentheses.}
\begin{tabular}{|llllll|}
\hline
$A_{1}$ & $k$ & Ref.[13] & $E_{o}$ & $E_{o}+E_{2}/\bar{l}^{2}$ &
 Eq.(11) \\
 \hline
0.4&0.0&0.894427191&0.894427191&0.894427191&0.894427191\\
  &0.01&0.9190495619&0.9199389105&0.9190185899&0.9190592557\\
  &0.05&0.99735023(19)&1.00292492& 0.99763976& 0.99732540\\
  &0.1 &1.0759666(05)& 1.08503730 & 1.07689328 & 1.07585618\\
  &0.2 &1.20488(59) & 1.21825 & 1.20662 & 1.20453 \\
  &0.3 &1.3138(20) & 1.32987 &1.31551 &1.31261\\
0.5&0.0&0.70710678119&0.70710678119&0.70710678119&0.70710678119\\
 &0.01&0.7174441845&0.7175816778&0.7174395344&0.7174446067\\
 &0.05&0.7548104279&0.7569796562&0.7546116940&0.7548645285\\
 &0.1&0.795714744277(07)&0.8013654436&0.7951332112&0.7959080738\\
 &0.2&0.86613531(67) &0.87841386& 0.86508153& 0.86648455\\
 &0.3&0.9269(68) & 0.9450 & 0.92586 & 0.92734\\
\hline
\end{tabular}
\end{center}
\end{table}
\newpage
\begin{table}
\begin{center}
\caption{Ground - state energies of a $\pi^{-}$ meson in
$V(r)=-A_{1}/r$ and $S(r)=kr$ (in $\hbar=c=m=1$ units).
The lower bounds to the
energies $E$ of Ref.[13] are obtained by replacing the last $j$ digits
of the upper bounds with the $j$ digits in parentheses.}
\begin{tabular}{|llllll|}
\hline
$A_{1}$ & $k$ & Ref.[13] & $E_{o}$ & $E_{o}+E_{2}/\bar{l}^{2}$ &
 Eq.(11) \\
 \hline
0.2&0.0&0.9789063129&0.9789063129&0.9789063129&0.9789063129\\
   &0.01&1.0266839(09) &1.02871502 &1.02708579 &1.02671248\\
   &0.05&1.145795(48) &1.1479170 &1.1449853 &1.1439393 \\
   &0.1 &1.263(34) &1.2543 &1.2506 &1.2491\\
   &0.2 &1.47(34) & 1.418& 1.413 &1.411 \\
   &0.3 &1.68(41) & 1.551 & 1.545 &1.542 \\
0.3&0.0&0.94868329805&0.94868329805&0.94868329805&0.948683298\\
   &0.01&0.9834119450(49)&0.9852632146&0.9835164748&0.9834118201\\
   &0.05&1.079797(62)&1.085772& 1.081002& 1.079875\\
   &0.1 &1.170(69) &1.17826 & 1.17182 & 1.16999\\
   &0.2 &1.316(05) &1.3227 & 1.31403 & 1.31133\\
   &0.3 &1.443(09) & 1.4404 & 1.42998 & 1.42666\\
\hline
\end{tabular}
\end{center}
\end{table}
\newpage
\begin{table}
\begin{center}
\caption{Ground - state energies of a $\pi^{-}$ meson in
$V(r)=-A_{1}/r$ and $S(r)=kr$ (in $\hbar=c=m=1$ units).
The lower bounds to the
energies $E$ of Ref.[13] are obtained by replacing the last $j$ digits
of the upper bounds with the $j$ digits in parentheses.}
\begin{tabular}{|llllll|}
\hline
$A_{1}$ & $k$ & Ref.[13] & $E_{o}$ & $E_{o}+E_{2}/\bar{l}^{2}$ &
 Eq.(11)\\
 \hline
0.4&0.0&0.894427191&0.894427191&0.894427191&0.894427191\\
  &0.01&0.9180779227&0.9190603882&0.9180537998&0.9180878824\\
  &0.05&0.9915050(49)&0.997801768&0.99204395&0.99150796\\
  &0.1 &1.063490(84) &1.07395616&1.06503089&1.06352424\\
  &0.2 &1.1791(88) &1.19492 &1.18212 & 1.17926\\
  &0.3 & 1.2755(37) &1.29449 & 1.27883 & 1.27503\\
0.5&0.0&0.7071067812&0.7071067812&0.7071067812&0.7071067812\\
  &0.01&0.7168151723&0.7169998899&0.7168096675&0.7168157252\\
  &0.05&0.7514351538&0.7544233036&0.7512342784&0.7514976377\\
  &0.1&0.78877520361(59)&0.7967081101& 0.7883364991& 0.7889588588\\
  &0.2&0.852303690(62)& 0.869749249& 0.852125792 & 0.852493296\\
  &0.3&0.9068223(16)&0.9322673 & 0.9074421 & 0.906889 \\
\hline
\end{tabular}
\end{center}
\end{table}
\end{document}